\documentclass[twocolumn,floatfix,aps,superscriptaddress,prb]{revtex4}
\usepackage{graphicx}
\usepackage{amsmath}
\usepackage[colorlinks=true, citecolor=blue, urlcolor=blue, linkcolor = blue]{hyperref}
\usepackage{amssymb}
\usepackage{bm}
\usepackage{color}
\usepackage{bookmark}
\usepackage{tabularx}
\usepackage{mathtools}
\usepackage{microtype}
\usepackage{cancel}
\usepackage{relsize}
\usepackage{mathrsfs}
\usepackage{ulem}

\usepackage{xcolor}

\begin{document}

\title{Classes of metastable Thermodynamic Quantum Time Crystals}
\author{S. I. Mukhin} 
\affiliation{Theoretical Physics and Quantum Technologies Department, NUST ``MISIS", Leninski Avenue 4, 119991 Moscow, Russia}
\author{T. R. Galimzyanov}

\affiliation{Theoretical Physics and Quantum Technologies Department, NUST ``MISIS", Leninski Avenue 4, 119991 Moscow, Russia}
\affiliation{A. N. Frumkin Institute of Physical Chemistry and Electrochemistry RAS, Moscow, Russia}

\date{}

\begin{abstract}
We found that thermodynamic quantum time crystals in Fermi systems, defined as quantum orders oscillating
periodically in the imaginary Matsubara time with zero mean, are metastable for two general classes of solutions.
Mean-field time independent solutions proved to have lower free energy manifesting true thermodynamic
equilibrium with either single or multiple (competing) charge, spin, and superconducting symmetry breaking
orders. The “no-go” theorem is proven analytically for a case of long-range interactions between fermions in
momentum space in electron-hole and Cooper channels.
\end{abstract}

\date{\today}
\maketitle


Mean-field  is widely used description for broken symmetry ordered states of many-body systems. Functional integral formalism relates the quantum partition function $Z$ of a many-body system to the Euclidean action $S$, where the inverse temperature, $\beta={1}/{T}$, acts as imaginary Matsubara time \cite{agd,hertz}:
\begin{align}
&Z=\int{\cal{D}}\bar{\Psi}(\tau){\cal{D}}\Psi (\tau)\exp{(-S)},\label{Z}\; \\
&S=\int^{\beta}_{0}d\tau\sum_{\vec{r}}\left\{\bar{\Psi}(\partial_{\tau}-\mu)\Psi+H(\bar{\Psi},\Psi)\right\},
\label{S0}
\end{align}
\noindent here $\bar{\Psi}, \Psi$ are Matsubara time conjugated quantum field operators and the Plank's constant is the unit of action ($\hbar=1$). It is also known that degenerate states with classically broken symmetries could be connected by instantons, i.e. by periodic in Matsubara time solutions of the classical equations of motion extremizing the Euclidean action \cite{1, harrington}.  One of us has proposed an example of such solution for the quasi one-dimensional onsite repulsive-U Hubbard model \cite{2,3,4}, introducing  a notion of {\it{quantum order parameter, QOP}}, that contains an "instantonic crystal" , which breaks translational invariance along the Matsubara time axis. It was demonstrated \cite{3,4} that instantonic crystal, e.g. of spin density wave type, possesses zero scattering cross-section for incident particles that couple to spin and, thus, forms a new kind of "hidden order", that could be relevant e.g. for  pseudo-gap state of high-Tc cuprates\cite{hidd}.  It was suggested recently \cite{efetov}, that in the case of two competing spin- and charge density wave orders an instantonic crystal, called there \cite{efetov} "thermodynamic quantum time crystal", might form and realize previously proposed "quantum time crystal"  \cite{wilczek} as the ground state. The purpose of the present paper is to prove analytically that at least for two general classes of solutions neither a single nor multiple (competing) charge-, spin- and superconducting "thermodynamic quantum time crystals" (instantonic crystals) can form a stable thermodynamic equilibrium state of interacting fermi-system at any temperature including the absolute zero. 
To start a proof of the 'no-go' theorem we consider a simple two band model with long-range interaction between electron-hole pairs in momentum space, that was recently considered in \cite{efetov}, and discussed previously \cite{3,4} in relation with quasi 1D model. Namely, we introduce a complex quantum order parameter, periodic in Matsubara time:
\begin{align}
\hat{H}_{M}=\begin{pmatrix}
    \epsilon_{q} +t_{q} & M(\tau)   \\
      M^*(\tau)&  - \epsilon_{q} +t_{q}
\end{pmatrix}
\label{lduv}
\end{align}
\noindent Hamiltonian $\hat{H}_{M}$ in Eq. (\ref{lduv}) acts on a two-component "spinor" of bare fermionic states assigned to each point in momentum space (Brillouin zone) , $\vec{\psi}^T\equiv \{u_{+},u_{-}\}$. Corresponding two energy bands possess dispersions counted from the chemical potential $\mu$: $\epsilon_{\pm}=\pm\epsilon_{q} +t_{q}$. The amplitudes $u_{\pm}$ of the fermion wave function could be either electron-  and hole- amplitudes in a two-band model \cite{efetov}, or amplitudes of "right"- and "left"- movers in a quasi-$1D$ model considered in  \cite{3,4}. A simple example of an origin of a complex field $M(\tau)\equiv M_{1}(\tau)+iM_{2}(\tau)$ is provided e.g. by a decoupling of onsite repulsion term in the Hubbard $U$ lattice model via Hubbard-Stratonovich (HS) procedure, that leads to the spin- and charge-density fields,  $M_1, M_2$ \cite{schulz}:
\begin{align}
&\exp{\left[-\int^{\beta}_{0}d\tau{U}\hat{n}_{i\uparrow}\hat{n}_{i\downarrow}\right]}=\frac{1}{\pi U}\int {\cal{D}}M_1{\cal{D}}M_2\nonumber\\
&\exp\left\{-\int^{\beta}_{0}d\tau \left[\frac{1}{U}(M_{1}^{2}+M_{2}^{2})+iM_2\hat{n}_{i}+M_1\hat{s}_{zi}\right]\right\},
\label{H}
\end{align}
\noindent where onsite-$i$ charge and spin density operators are defined as:
\begin{align}
\hat{n}_{i}=\hat{n}_{i\uparrow}+\hat{n}_{i\downarrow},\;\hat{s}_{zi}=\hat{n}_{i\uparrow}-\hat{n}_{i\downarrow}.
\label{def}
\end{align}
\noindent
Hence, partition function in Eqs. (\ref{Z}), (\ref{S0}), after HS decoupling in Eq. (\ref{H}) and changing from lattice coordinate to momentum representation, is expressed as follows (a single $U$ is generalized by a different $U_{\gamma}$ for two fields $M_{\gamma=1,2}$):
\begin{align}
&Z=\displaystyle\int{\cal{D}}M_1{\cal{D}}M_2\prod_q{\cal{D}}\bar{\psi}_q(\tau){\cal{D}}\psi_q(\tau) \exp{\left\{-S\right\}}\label{euclid}\\
&S=\displaystyle\int_0^{\beta}d\tau \sum_q \left\{\bar{\psi}_q\left[\partial_{\tau}+\hat{H}_{M}\right]\psi_q+\sum_{\gamma=1}^{2}\dfrac{M_{\gamma}^2(\tau)}{4U_{\gamma}}\right\}
\label{S}
\end{align}

\noindent In Eqs. (\ref{euclid}), (\ref{S}) path integration implements a trace over diagonal elements of the exponential, and hence it is performed over $\tau$-periodic Hubbard-Stratonovich fields:
\begin{eqnarray}
{M}_{\gamma}\left(\tau+{1}/{T}\right)={M}_{\gamma}(\tau),
\label{SHM}
\end{eqnarray}
\noindent In case of a classical phase transition the overwhelming contribution to the path integral $Z$ comes from the $\tau$-independent HS fields, i.e. the minimum of the Euclidian action $S$ is achieved with some particular $\tau$-independent functions $M_{\gamma}(i)$, that constitute well known classical (mean-field) order parameters (COP). A condition for  the minimum of $S$ , from which the COP is found, is called the self-consistency mean-field equation, and was first introduced by P. Weiss \cite{weiss} for ferromagnetic domains. It was shown \cite{2,3,4} that besides COP, there exist other minima of the Euclidian action $S$, described with Hubbard-Stratonovich fields, that were called {\it{quantum order parameters}} (QOP), being $\tau$-periodic functions with zero mean:
\begin{eqnarray}
\langle M(\tau)\rangle_{1/T}=0.
\label{me}
\end{eqnarray}
 \noindent 
The required integration over Grassmann fields $\bar{\psi}_q(\tau),\, \psi_q(\tau)$ in the partition function $Z$, (\ref{euclid}), leads to the functional determinant $Det[\partial_{\tau}+\hat{H}_{M}]=\prod_m \epsilon_m(M)$, where eigenvalues $\epsilon(M)$ with the corresponding fermionic eigenvectors $\vec{\xi}(\tau)$ are defined as  \cite{neveu,2}:

\begin{eqnarray}
(\partial_{\tau}+\hat{H}_{M})\vec{\xi}_m=\epsilon_m\vec{\xi}_m;\;\vec{\xi}_m(\tau+{1}/{T})=-\vec{\xi}_m(\tau)
\label{fs}
\end{eqnarray}
\noindent The eigenvalues $\epsilon_m$ could be obtained using the following procedure. First, a spectrum $\{\alpha_q\}$ of the quasi-energies (Floquet indices) of the  Matsubara time-dependent Hamiltonian Eq. (\ref{lduv}) is found \cite{neveu, 2}. The Floquet indices label 'Bloch' solutions of the corresponding Dirac like equation with $\tau$-periodic potential $ M(\tau)$ in (\ref{lduv}):
\begin{eqnarray}
&(\partial_{\tau}+\hat{H}_{M})\vec{\psi}_q=0, 
\label{bloch}
&\vec{{\psi}}_q(\tau+{1}/{T})=e^{-\alpha_q}\vec{\psi}_q(\tau).
\label{FQF}
\end{eqnarray}
\noindent Provided the 'Bloch' functions $\vec{\psi}_q$ and indices $\alpha_q$ are known, the eigenvalues follow: $\epsilon_{m,q}=i(2m+1)\pi/T+\alpha_q$, enabling antiperiodicity of the fermionic eigenfunctions in (\ref{fs}), that are constructed as: $\vec{\xi}_{m,q}=\exp\{i(2m+1)\pi \tau/T+\alpha_q\tau/T\}\vec{\psi}_q$. Calculating product $\prod_{m,q}\epsilon_{m,q}$ in  the partition function in Eq. (\ref{euclid}) one finds \cite{neveu,2}:
\begin{eqnarray}
Z=\int{\cal{D}}{M(\tau}) exp{\{-S_{M}\}}\prod_q \cosh\left(\dfrac{\alpha_q}{2}\right)
\label{zm}
\end{eqnarray}
\noindent where $S_{M}$ is the bare Gaussian action of the Hubbard-Stratonovich fields $M_{\gamma}(\tau)$ expressed by the last sum in Eq. (\ref{S}).
\noindent Then, QOP is a periodic function $M(\tau)$ that obeys Eqs. (\ref{SHM}), (\ref{me}) and extremizes the total action, being a saddle-point of the path integral (\ref{zm}) in the functional space of HS fields: 
\begin{align}
\delta_{M(\tau)}{\left\{ S_M-\sum_q ln\left\{\cosh\left(\dfrac{\alpha_q}{2}\right)\right\}\right\}}=0
\label{self}
\end{align}
\noindent  The self-consistency equation is readily derived from (\ref{self}) using the first-order perturbation theory\cite{neveu}:
\begin{align}
{T\partial_{M(\tau)}\alpha_q={\bar{\psi}}_q(\tau)\{\partial_{M(\tau)}\hat{H}_{M}\}{\psi}_q(\tau)}
\label{DFQ}
\end{align} 
\noindent where a normalization condition is assumed:
\begin{align}
\int_0^{\beta}d\tau{\bar{\psi}}_q(\tau){\psi}_q(\tau)=1
\label{nrmlz}
\end{align}
\noindent Substituting this condition and Eq. (\ref{DFQ}) into Eq. (\ref{self}) one obtains:
\begin{align}
\dfrac{M(\tau)}{U}= \sum_q \tanh\left(\dfrac{\alpha_q}{2}\right){\bar{\psi}}_q(\tau)\{\partial_{M(\tau)}\hat{H}_{M}\}{\psi}_q(\tau)
\label{fins}
\end{align}
\noindent Rather nontrivial functional equation Eq. (\ref{fins}), where we have dropped the indices $\gamma=1,2$ of $M(\tau)$, has to be solved in each point of the Matsubara time interval $[0,1/T]$.

\noindent  Resuming consideration of the Dirac-type equation (\ref{bloch}) with Hamiltonian matrix given by Eq. (\ref{lduv}) we gauge out the 'antinesting' part of the dispersion, $t_{q}$, by "rotation" in Matsubara time: 
\begin{eqnarray}
\vec{\psi}\equiv 
 \left(\begin{array}{c}
u_{+} \\
u_{-}
\end{array}\right)=
  e^{-\tau t_{q}}\vec{\phi}\equiv  e^{-\tau t_{q}}\left(\begin{array}{c}
g_{+}\\
g_{-}
\end{array}\right)
\label{ganest}
\end{eqnarray}
\noindent After that the famous "nesting" symmetry \cite{keldysh} $\epsilon_{+}= -\epsilon_{-}$ is restored in the resulting Dirac-type equation:
\begin{align}
\begin{pmatrix}
    \partial_{\tau}+\epsilon_{q}  & M(\tau)   \\
      M^*(\tau)&   \partial_{\tau}- \epsilon_{q} 
\end{pmatrix}\vec{\phi}_q(\tau)=0
\label{tnest}
\end{align}

\noindent Thus, Floquet indices in the representation of 'rotated'  spinor $\vec{\phi}_q(\tau)$ become shifted:
\begin{eqnarray}
\alpha_{q}=T^{-1}t_{q}+\tilde{\alpha}_{q};\;\;\;\vec{\phi}_q(\tau+1/T)=e^{-\tilde{\alpha}_{q}}\vec{\phi}_q(\tau).
\label{floqq}
\end{eqnarray}
\noindent We shall see below from exact solution, that the indices $\tilde{\alpha}_{q}$  occur in plus-minus pairs, obeying the symmetry relation:
\begin{eqnarray}
\tilde{\alpha}(-\epsilon_{q})=-\tilde{\alpha}(\epsilon_{q})
\label{conj}
\end{eqnarray}
\noindent Next, we  imply an electron-hole symmetry of the bare spectrum $\epsilon_q$ and combine it with the symmetry relation (\ref{conj}). As a result, in the representation of  $\tilde{\alpha}_{q}$, Euclidean action from Eq. (\ref{zm}) acquires the following form allowing for identity $\cosh[(x+y)/2]\cosh[(x-y)/2]=(\cosh(x)+\cosh(y))/2$:
\begin{eqnarray}
&S^Q=-\frac{1}{2}\displaystyle\sum_q\ln\frac{1}{2}[\cosh(\tilde{\alpha}_{q})+\nonumber\\
&+\cosh(t_{q}/T)]+\displaystyle\sum_{\gamma=1}^{2}\dfrac{{P_\gamma}^2}{4\tilde{U}_{\gamma}}\label{smn}
\end{eqnarray}
\noindent  where $\tilde{U}_{\gamma}$ stands for parameters ${U}_{\gamma}$ in (\ref{S}), but properly renormalized by the volume of the system, that is tacitly involved in the summation over momenta $\sum_q$ over the fermionic states in the Brillouin zone. Here we have also introduced notations for 'mean square orders':
\begin{eqnarray} 
{P_\gamma}^{2}= T\int\limits_0^{1/T} {M_{\gamma}^2(\tau) } d\tau 
\label{mmm}
\end{eqnarray}
\noindent
First,  we are going to prove the 'no-go' theorem for the action (\ref{smn}) for two general classes of HS fields: 1) $M_2(\tau)=$const, and 2) $M_1(\tau)+iM_2(\tau)\equiv M(\tau)e^{i\phi}$ with $\phi=$const and $M(\tau)$ real function, see Fig.\ref{nogo12} :
\begin{figure}[h!!]
\centerline{\includegraphics[width=1.\linewidth]{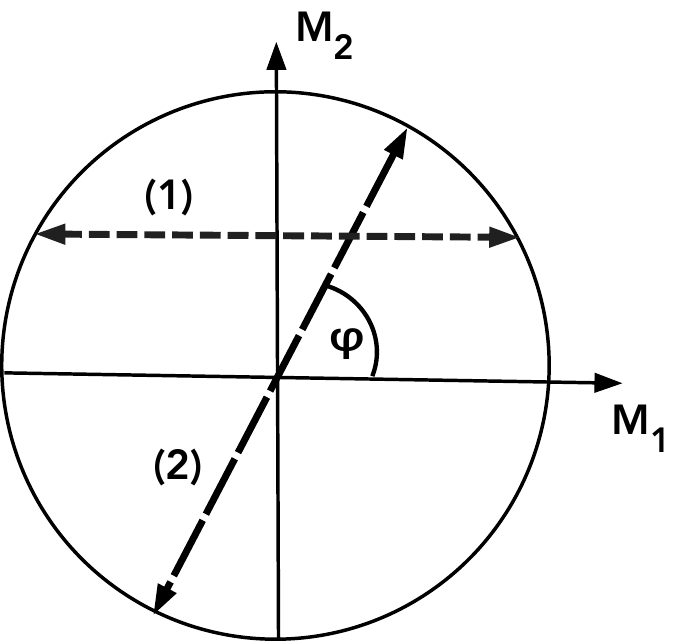}}
\caption{Schematic layout of QOP variation in Matsubara time. Chord (1), short dashes, corresponds to: $M_1(\tau+1/T)=M_1(\tau)$, $M_2(\tau)=$const. Chord (2), long dashes, corresponds to: $M_1(\tau)+iM_2(\tau)\equiv M(\tau)e^{i\phi}$, with $\phi=$const.}
\label{nogo12}
\end{figure}
\noindent Namely, we are going to demonstrate that in the both cases Euclidean action $S$ in (\ref{smn}) achieves its minimum only under condition $M_1(\tau)=$const, $M_2(\tau)=$const, i.e. only COP could be its thermodynamic equilibrium state at any temperature including $T=0$. The
case in \cite{efetov} corresponds to $\phi=\phi(\tau)\neq$const and, therefore, does not belong to the classes of HS fields that we consider in this work.

\noindent Substituting constant $M_\gamma(\tau) =const=P^{c}_\gamma$ into Eq. (\ref{tnest}) and using definitions (\ref{floqq}) we find Floquet indices (i.e. spectrum) of the fermi-system with COP: $T\alpha_q=t_q\pm\sqrt{\sum_\gamma {P^{c2}_\gamma}+\epsilon_q^2}$. This leads to the Euclidean action:
\begin{eqnarray}
&&S^{C}=-\frac{1}{2}\sum_q\ln{\frac{1}{2}[\cosh\left(\frac{\sqrt{\sum_\gamma {P^{c2}_\gamma}+\epsilon_q^2}}{T}\right)}+ \nonumber\\
&&{ \cosh({t_{q}}/{T})]}+\displaystyle\sum_{\gamma=1}^{2}\dfrac{{P^{c2}_\gamma}}{4\tilde{U}_{\gamma}}
\label{scop}
\end{eqnarray}
\noindent A direct comparison of expressions (\ref{smn}) and (\ref{scop}) indicates that for any COP and QOP  states with equal 'mean square orders' $P^{c2}_\gamma$ and ${P_\gamma}^2$ the difference between their Euclidean actions (e.g. free energies) depends merely on the difference between their Floquet spectra $\tilde{\alpha}_{q}$ and  $\sqrt{\sum_\gamma {P^{c2}_\gamma}+\epsilon_q^2}/T$. The key of our proof is the demonstration that any Matsubara time-dependent order parameter QOP has Floquet spectrum $\tilde{\alpha}_q  < \sqrt{\sum_\gamma {P^{c2}_\gamma}+\epsilon_q^2}/T$ for all $q$. Hence, corresponding QOP free energy is greater than that of COP and, consequently, the QOP state is metastable.

Consider case 1) corresponding to $M_2=$const and $M_1(\tau)$ periodic with period  $1/T$ along the Matsubara time axis. After an extra unitary transformation:
\begin{eqnarray}
f_{\pm}=(2)^{-1/2}({g}_+\pm {g}_-) 
\label{efs}
\end{eqnarray}
\noindent
the corresponding Floquet equation for the 'spinor' $\vec{\phi}(\tau)$ defined in Eq. (\ref{ganest}) is readily obtained from  (\ref{tnest}), allowing for the definition (\ref{mmm}):
\begin{eqnarray}
\label{ff}
&&({\partial^2}_{\tau}-Q_{\pm}(\tau)-\epsilon_{q}^2-{P_2}^{2})f_{\pm}=0;\label{dbdgf}\;\\
&&Q_{\pm}(\tau)=M_1(\tau)^2\mp \partial_{\tau}M_1(\tau)
\label{dbdg}
\end{eqnarray} 
\noindent  Rewriting equation (\ref{dbdgf})  in the following equivalent form and integrating over one period along the Matsubara time axis, 1/T, we obtain:
\begin{eqnarray}
T\int\limits_0^{1/T} {\frac{{\partial _\tau ^2 f_ \pm  (\tau )}}{{f_ \pm  (\tau )}}d\tau }  = T\int\limits_0^{1/T} {\left[ {Q_ \pm  (\tau ) + \epsilon _q^2+{P_2}^{2} } \right]} d\tau 
\label{fratioi}
\end{eqnarray}
\noindent Right hand side of Eq. (\ref{fratioi}) can be simplified using definitions of $Q(\tau)$ in Eq. (\ref{dbdg}) and  ${P_\gamma}^{2}$ in Eq. (\ref{mmm}), together with periodicity condition (\ref{SHM}). Simultaneously, the left hand side of Eq. (\ref{fratioi}) could be directly expressed via $\tilde{\alpha}_{q}$ using Eq. (\ref{floqq}) in the form: $f_{\pm}(\tau)=e^{-\tilde{\alpha}_{q}\tau T}\theta_{\pm}(\tau)$ with $\theta_{\pm}(\tau+1/T)=\theta_{\pm}(\tau)$. Thus, after straightforward manipulations, Eq. (\ref{fratioi}) acquires an equivalent form:
\begin{eqnarray}
&\displaystyle T\int\limits_0^{1/T} {\frac{{\partial _\tau ^2 f_ \pm  (\tau )}}{{f_ \pm  (\tau )}}d\tau }\equiv(\tilde{\alpha}_{q} T)^2  - 2\tilde{\alpha}_{q}T^2 \int\limits_0^{1/T} {\frac{{\dot \theta_{\pm}  (\tau )}}{{\theta_{\pm}  (\tau )}}d\tau } \nonumber \\
& +\displaystyle T\int\limits_0^{1/T} {\frac{{\ddot \theta_{\pm}  (\tau )}}{{\theta_{\pm}  (\tau )}}d\tau } =\epsilon _q^2+\sum_\alpha {P_\alpha}^{2}.
\label{fratioi1}
\end{eqnarray}
\noindent Integrating by parts in the left hand side of (\ref{fratioi1}) and allowing for the periodicity of function $\theta_{\pm}(\tau )$ one finds:
\begin{eqnarray}
&(\tilde{\alpha}_{q} T)^2=-T\displaystyle\int\limits_0^{1/T} \frac{{\dot \theta_{\pm} ^2 (\tau )}}{{\theta_{\pm} ^2 (\tau )}}d\tau + \epsilon _q^2+\sum_\alpha {P_\alpha}^{2}\nonumber\\
&\leq \epsilon _q^2+\sum_\alpha {P_\alpha}
\label{fratioi2}
\end{eqnarray}
\noindent Since both the Floquet indices and Bloch functions in Eqs. (\ref{dbdgf}) are real (see below), the equality in Eq. (\ref{fratioi2}) could be achieved only in the COP case:
\begin{eqnarray}
T\int\limits_0^{1/T} \frac{{\dot \theta_{\pm} ^2 (\tau )}}{{\theta_{\pm} ^2 (\tau )}}d\tau=0;\;\theta_{\pm}(\tau )\equiv \text{const}
\label{COP}
\end{eqnarray}
\noindent Hence, we had proven that Euclidean action (free energy) of any QOP (thermodynamic quantum time crystal) state, $S^Q$, is always higher than that of the COP state, $S^C$, and therefore, QOP state is metastable. 

 A proof for the second class of thermodynamic quantum time crystals, i.e.: $M_1(\tau)+iM_2(\tau)\equiv M(\tau)e^{i\phi}$ with $\phi=$const and $M(\tau)$ real function, is trivially reduced to the case 1 considered above by the following transformation of  the 'spinor' $\vec{\phi}_q(\tau)$ in Eq. (\ref{tnest}):
\begin{align}
\vec{\phi}_q(\tau)\equiv
\begin{pmatrix}
   {g}_{+} \\
    {g}_{-}  
\end{pmatrix}=
\begin{pmatrix}
   e^{i\phi/2}\tilde{g}_{+} \\
    e^{-i\phi/2}\tilde{g}_{-}  
\end{pmatrix}
\label{gg}
\end{align}
\noindent Then, Eq. (\ref{tnest}) with $M(\tau)e^{\pm i\phi}$ functions in the place of  functions $M(\tau)$, $M^*(\tau)$,  transforms into the following equation: 
\begin{align}
\begin{pmatrix}
    \partial_{\tau}+\epsilon_{q}  &M(\tau) \\
      M(\tau)&   \partial_{\tau}- \epsilon_{q} 
\end{pmatrix}\vec{\tilde{\phi}}_q(\tau)=0
\label{tnest1}
\end{align}
\noindent with real function $M(\tau)$. Here $(\vec{\tilde{\phi}})^T\equiv \{\tilde{g}_{+},\tilde{g}_{-}\}$. Now, substituting $M_1(\tau)\rightarrow M(\tau) $ and $M_2(\tau)\equiv 0$ everywhere in the above proof in case 1, we arrive as well at the proof of the statement $S^Q>S^C$ also for the case 2 considered in this section.
Next, introducing also superconducting order parameter $\Delta_q$ of a d-wave symmetric type in momentum space  (e.g. relevant for high-T$_c$ cuprates) we prove metastability of  thermodynamic quantum time crystals also in the case of multiple (competing) charge, spin and superconducting symmetry breaking orders described by the $4\times4$ matrix in the 'bispinor' space \cite{matmuk} $\vec{\Psi}^T\equiv\{u_+,u_-,v_+,v_-\}$:

\begin{align}
  \partial_{\tau} \vec{\Psi}_q+
\begin{pmatrix}\epsilon_{q} &M& \Delta_{q} &0 \\
M^*& -\epsilon_{q} & 0& -\Delta_{q}\\
 \Delta_{q}^*  &0 & -\epsilon_{q}& M \\
 0 & -\Delta_{q}^* & M^*&\epsilon_{q}
\end{pmatrix}\vec{\Psi}_q =0
\label{cmd}
\end{align}
\noindent It is straightforward now to reduce Hamiltonian matrix in (\ref{cmd}) to two block-matrices $2\times2$ of the kind (\ref{tnest}) by imposing the following linear relations between bispinor components \cite{matmuk} : $v_{\pm}=\gamma_\pm u_\mp$ with constant coefficients: 
\begin{align}
|\gamma_\pm|=1;\;\gamma_-=-\gamma_+^*\frac{\Delta^*}{\Delta}
\label{lins}
\end{align}
\noindent This reduces Eq. (\ref{cmd}) to the two equations of the kind Eq. (\ref{tnest}), but with different composite order parameters $M_\pm$:
\begin{align}
M_+= M+\gamma_+\Delta; \; M_-=(M\gamma_-/\gamma_++\Delta^*/\gamma_+)^*
\label{mms}
\end{align}
\noindent Hence, our proof of the metastability of the thermodynamic quantum time crystals presented above applies also in the cases of coexisting spin-, charge and superconducting orders entering Hamiltonian in Eq. (\ref{cmd}). Depending on what condition of the considered  cases 1) or 2)  apply to the composite complex order parameters $M_\pm$ in Eq. (\ref{mms}), the corresponding version of the 'no-go' theorem stays.

Finally, we discuss analytic thermodynamic quantum time crystal solution obtained by one of us earlier \cite{2,3,4}, that provides direct  demonstration of the workings of 'no-go' theorem presented above and a 'pseudo-gap' thermodynamic behaviour. 
Comparing Eqs.(\ref{dbdgf}) and (\ref{dbdg}) with the well known solitonic-lattice equations for the Peierls/polyacetylene \cite{ssh,br,maki,machida} and one-dimensional Hubbard model \cite{matmuk} we conclude, that due to replacement of the space coordinate with imaginary Matsubara's time $\tau$ equation (\ref{dbdgf})  differs from the solitonic-lattice equations only by the opposite sign in front of the square dispersion $\epsilon_{q}^2$. Using this similarity one finds \cite{2,3,4} QOP  $M_1(\tau)$, that obeys self-consistency equation (\ref{fins}):
\begin{align}
M_1(\tau)=4nKTk_1 sn\left(4nKT\tau;k_1\right),\,K=K(k_1)
\label{norder}
\end{align}
\noindent Here $sn(\tau,k_1)$ is the Jacobi snoidal elliptic function, with period $1/nT$  commensurate with the main period $1/T$: $M_1(\tau)=M_1(\tau+1/nT)$,  integer $n=1,2,...$ counts number of instanton - anti-instanton pairs.  For simplicity, we consider here the case 1) of 'no-go' theorem with $M_2\equiv 0$.  
\begin{figure}[h!!]
\centerline{\includegraphics[width=1.\linewidth]{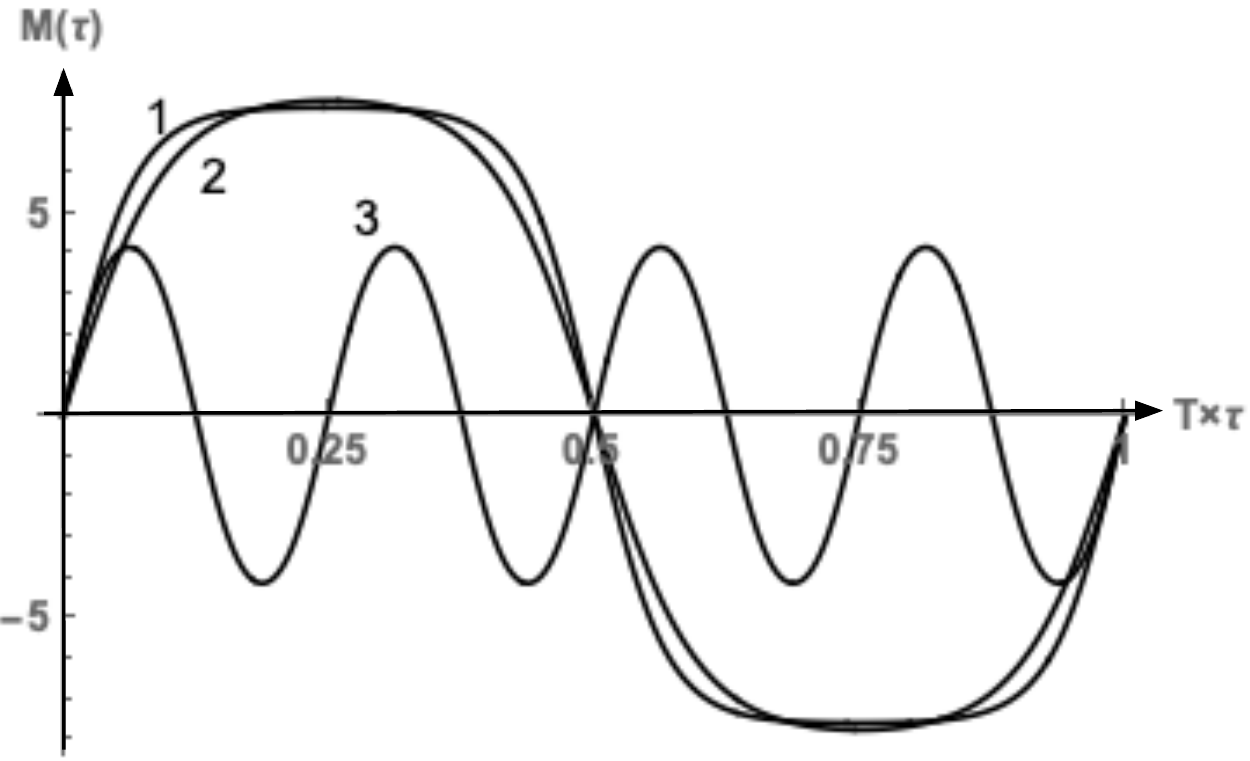}}
\caption{QOP function $M_1(\tau)$ as function of Matsubara time $\tau$. Curves are marked according to different parameter sets: (1) T=0.51; k=0.99999; n=1; (2) T=0.8; k=0.999; n=1; (3) T=0.51, k=0.84, n=4. Temperature $T$ is measured in arbitrary units and parameter $k$ is Landen transformed parameter $k_1$.}
\label{QOPM}
\end{figure}
\noindent Simultaneously, the Floquet indices spectrum $\tilde{\alpha}_{q}$ takes the form \cite{2} :
\begin{align}
\tilde{\alpha}_q=2\tilde{\epsilon}_q\left(\dfrac{1-{{k}}^2+\tilde{\epsilon}_q^2}{1+\tilde{\epsilon}_q^2}\right)^{1/2}
n\Pi\left(\dfrac{k^2}{1+\tilde{\epsilon}_q^2},{k}\right)
\label{flon}
\end{align}
\noindent  Accordingly, $\Pi(m,k)$ and $K(k_1)$ are elliptic integrals of the third and first kind respectively, and $k$ is Landen transformed parameter $k_1$ from (\ref{norder}): 
\begin{align}
&\tilde{\epsilon}_q\equiv \dfrac{\epsilon_{q}}{2TnK(k)},\;{k}=2\sqrt{k_1}/(1+k_1);\,k'^2={1-k^2}
\label{renorm}
\end{align}
\begin{figure}[h!!]
\centerline{\includegraphics[width=1.\linewidth]{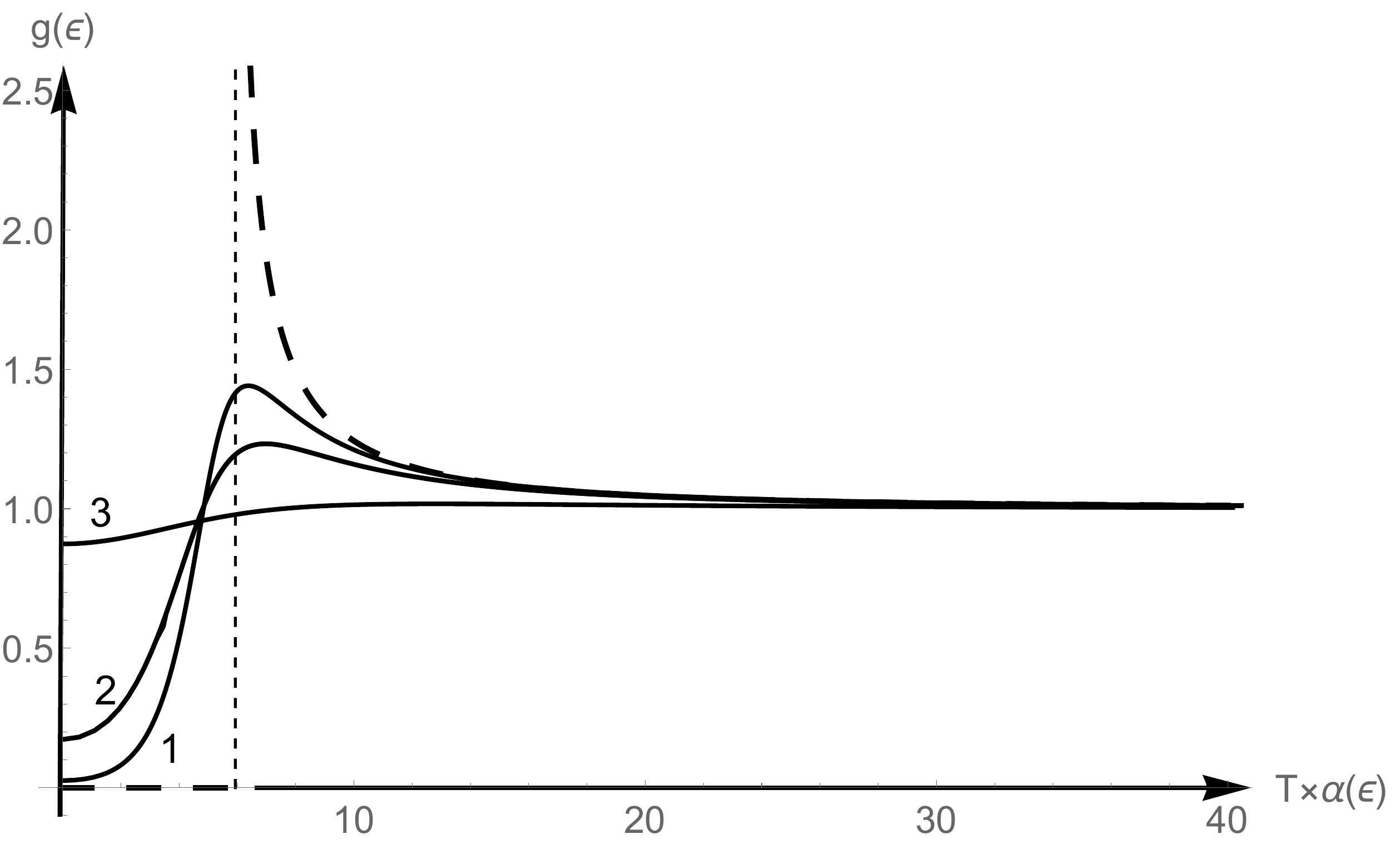}}
\caption{Density of Floquet states $\tilde{\alpha}_q$ (solid lines) for the same parameter sets (1)-(3) as indicated in Fig. \ref{QOPM}. Dashed lines is COP density of states  (\ref{scop}) with dispersion $\pm\sqrt{P^{c2}_1+\epsilon_q^2}$ under the equality condition (\ref{pgs}) between  COP and QOP  'mean square orders'  defined in (\ref{mmm}).}
\label{PSG}
\end{figure}
\noindent Hence, (\ref{flon}) proves the symmetry relation (\ref{conj}). The Jacobi function from Eq. (\ref{norder}) transforms the QOP self-consistency 
equation (\ref{fins}) into algebraic equation for parameters $k,n$:
\begin{align}
&\sum_{\vec{q}} \left[\tanh\dfrac{{\alpha}_q}{2}\right]\dfrac{\epsilon_{\vec{q}}}{\{(\epsilon^{2}_{\vec{q}}+ {\Delta^2}_{T})(\epsilon^{2}_{\vec{q}}+
k'^2{\Delta^2}_{T})\}^{1/2}}=\dfrac{1}{U}
\label{selfinn}\\
&\alpha_{\vec{q}}=T^{-1}t_{\vec{q}}+\tilde{\alpha}_{\vec{q}},\;\Delta_{T}\equiv 2TnK(k)
\label{tqop}
\end{align}

\noindent Result (\ref{flon}) is remarkable. Namely, compare expressions (\ref{smn}) and (\ref{scop}) under condition \cite{3}:

 \begin{align}
P^{c2}_1=P_1^2\equiv(2nK(k)T)^2\left(1+k'^2-2\dfrac{E(k)}{K(k)}\right),
\label{pgs}
\end{align}
\noindent while: $ P^{c2}_2=P_2^2=0$. Then, partition function in (\ref{smn}) maps the QOP state on the fermi gas with effective dispersion  $\varepsilon_{eff}(q)\equiv T\tilde{\alpha}_{\vec{q}}$ of the manifestly pseudo-gap type, while (\ref{scop}) possesses dispersion $\varepsilon_{eff}(q)=\pm\sqrt{P^{c2}_1+\epsilon_q^2}$ with the usual Peierls-type gap in the density of states $g$. Comparison of the corresponding densities of states $g(\varepsilon_{eff}(q))=(\partial \varepsilon_{eff}(q)/\partial \epsilon_q)^{-1}$ in Fig. \ref{PSG} makes then our 'no-go' theorem rather obvious, as the energy gain of the fermions in the gapped state is manifestly greater than in the pseudo-gap state at the same temperature.
Simultaneously, a comparison between the curves enumerated  as (1)-(3) in Figs. \ref{QOPM},\ref{PSG} indicates that instantonic crystals/thermodynamic quantum time crystals with more rectangular shape (i.e. $k\rightarrow 1$) of Jacobi snoidal function (\ref{norder}), see curves (1),(2) in Fig.\ref{QOPM}, create deeper pseudogap in the density of the Floquet states in Fig. \ref{PSG}, while  instantonic crystals of sine-like shape, see curve (3) in Fig. \ref{QOPM}, create shallow pseudogap in the density of the Floquet states according to curve (3) in Fig. \ref{PSG}. 


The authors acknowledge useful discussions with Serguey Brazovskii, Jan Zaanen and Konstantin Efetov. This research was supported by the Ministry of Science and Higher Education of the Russian Federation in the framework of Increase Competitiveness Program of NUST "MISiS" grant                 K2-2017-085, and via 'Goszadaniye' grant 3.3360.2017/PH.

\end{document}